\documentclass[12pt,preprint]{aastex}

\def\h13cop{H$^{13}$CO$^+$}
\def\hc5n{HC$_{5}$N}

\def\t32{$J=3-2$}
\def\h13cn{H$^{13}$CN}
\def\cc34s{CC$^{34}$S}
\def\n2hp{N$_2$H$^+$}

\def\13co{$^{13}$CO}
\def\c18o{C$^{18}$O}
\def\ch3cn{CH$_{3}$CN}
\def\c34s{C$^{34}$S}
\def\3423{$3_4-2_3$}

\def\deg{\hbox{$^{\circ}$}}

\def\arcsec{\hbox{$^{\prime\prime}$}}

\def\cc{cm$^{-3}$}

\def\h2{$H_{2}$}

\slugcomment{Updated on 2012/03/13, accepted by the ApJ Letters}

\shorttitle{Magnetic Field in The Massive Dense Clump IRAS 20126+4104}
\shortauthors{Shinnaga et al.}

\begin{document}


\title{Magnetic Field in The Isolated Massive Dense Clump IRAS 20126+4104}


\author{Hiroko Shinnaga,\altaffilmark{1} Giles Novak,\altaffilmark{2}  John E. Vaillancourt,\altaffilmark{3}  Masahiro N. Machida,\altaffilmark{4}  Akimasa Kataoka,\altaffilmark{5} Kohji Tomisaka,\altaffilmark{6} Jacqueline Davidson,\altaffilmark{7}  
Thomas G. Phillips,\altaffilmark{1,8} C. Darren Dowell,\altaffilmark{9} Lerothodi Leeuw,\altaffilmark{10,11}  
and Martin Houde\altaffilmark{12,7}  
}
\altaffiltext{1}{California Institute of Technology Submillimeter Observatory, 111 Nowelo St. Hilo HI 96720 U.S.A.}
\altaffiltext{2}{Northwestern University, 633 Clark Street  Evanston, IL 60208 U.S.A.}
\altaffiltext{3}{Stratospheric Observatory for Infrared Astronomy, Universities Space Research Association, NASA Ames Research Center, Moffet Field, CA 94035, U.S.A. }
\altaffiltext{4}{Department of Earth and Planetary Sciences, Faculty of Sciences, Kyushu University, 6-10-1 Hakozaki, Higashi-ku, Fukuoka 812-8581, Japan}
\altaffiltext{5}{Department of Astronomy, Kyoto University, Kitashirakawa-Oiwake-cho, Sakyo-ku, Kyoto 606-8502, Japan}
\altaffiltext{6}{National Astronomical Observatory of Japan and Department of Astronomy, School of Physical Sciences, Graduate University for Advanced Studies (SOKENDAI), Osawa 2-21-1, Mitaka, Tokyo 181-8588, Japan}
\altaffiltext{7}{University of Western Australia, 35 Stirling Highway, Crawley WA 6009 Perth, Australia}
\altaffiltext{8}{Division of Physics, Mathematics and Astronomy, California Institute of Technology,  MC 320-47, Pasadena, CA 91125 U.S.A.}
\altaffiltext{9}{Jet Propulsion Laboratory, California Institute of Technology, MS 169-506, 4800 Oak Grove Drive, Pasadena, CA 91109 U.S.A.}
\altaffiltext{10}{SETI Institute, 515 N. Whisman Avenue, Mountain View, CA, 94043, U.S.A.}
\altaffiltext{11}{Physics Department, University of Johannesburg, P.O. Box 524, Auckland Park, 2006, South Africa}
\altaffiltext{12}{University of Western Ontario, 1151 Richmond Street, London, Ontario, Canada, N6A 3K7 Canada}


\begin{abstract}
We measured polarized dust emission at 350$\mu$m towards the high-mass star forming massive dense clump 
IRAS 20126+4104 
using the SHARC II Polarimeter, SHARP, at the Caltech Submillimeter Observatory. 
Most of the observed magnetic field vectors agree well with magnetic field vectors obtained from 
a numerical simulation 
for the case when the global magnetic field lines are inclined with respect to the rotation axis of the dense clump. 
The results of the numerical simulation show that rotation plays an important role 
on the evolution of the massive dense clump and its magnetic field.  
The direction of the cold CO 1$-$0 bipolar outflow is parallel to the observed magnetic field within the dense clump 
as well as the global magnetic field, as inferred from optical polarimetry data, 
indicating that the magnetic field also plays a critical role in an early stage of massive star formation.  
The large-scale Keplerian disk of the massive (proto)star rotates in almost 
opposite sense to the clump's envelope.  
The observed magnetic field morphology and the counter-rotating feature of the massive dense clump 
system provide 
hints to constrain the role of magnetic fields in the process of high mass star formation.  
\end{abstract}






\section{Introduction}
Massive stars deposit 
energy into the interstellar medium, 
hence they may play a key role in regulating star formation in galaxies.  
Nevertheless, 
massive star formation is still poorly understood.  

There are many difficulties in studying the formation of massive stars.  
Massive stars are rare and 
evolve quickly. 
The youngest phases of massive stars is poorly understood because 
they are deeply embedded in dense clumps. 
Often times massive stars form in cluster regions, which 
makes it difficult to disentangle the activities of each young massive star.  

The early B type massive (proto)star, IRAS 20126+4104 ($\sim$ 7 M$_{\odot}$), is a well studied 
unique target that allows us to observe the early stages of massive star formation in a simple configuration.    
The natal massive dense clump, which is a site of massive star formation,  
is an isolated rotating (2 kms$^{-1}$pc$^{-1}$) clump \citep{shi08} with a mass of $\sim$ 200 M$_{\odot}$ and temperature of 40 K \citep{she00}.  It has a large scale bipolar outflow with a size scale of 0.5 pc 
\citep{she00,shi08}.  
The direction of this flow 
is very different from the direction of the smaller scale jet emanating from the massive (proto)star \citep{ces99}, indicating that the jet might be precessing.   
The radial column density profile of the massive dense clump shows a shallow slope (r$^{-0.2}$) in an inner region 
within a radius of $\sim$ 0.13 pc, while the outer region of radius $\ga$ 0.13 pc has a steeper slope ($\sim$ r$^{-1.3}$),   
indicating that 
the inner region may be experiencing infall, while the infalling wave has not yet reached the outer region \citep{shi08}.  
Another interesting feature of the massive dense clump is that 
the Keplerian  disk, known to be associated with the (proto)star \citep{ces99,ces05}, 
rotates almost in an opposite sense with respect to the rotation of the massive dense clump \citep{shi08}.  

Magnetic fields may play critical roles in star formation \citep{mck07}. 
We here report a study of the magnetic field of the massive dense clump, IRAS 20126+4104.  

\section{Observations and Data Reduction}
We used the SHARC II Polarimeter, SHARP \citep{li08}, with the 10.4 meter Leighton telescope 
at the Caltech Submillimeter Observatory (CSO) to measure dust polarization in the massive dense clump 
at 350 $\mu$m during 2007 August.  
SHARP uses the detector in SHARC II \citep{dow03} which  
has 384 pixels. 
Using the optics of SHARP, 
one divides the detector array into two sections, referred to as the H and V subarrays,  
for two orthogonal polarization components.  
Both polarization components are thus observed simultaneously.  
Each H and V subarray has 12 $\times$ 12 pixels with a 
field of view of $\sim$ 1' $\times$ 1'.  
The beam size was measured to be 9\arcsec.  
The data were acquired under good weather conditions.  The 225 GHz opacity was 
measured to be from 0.04 to 0.07.   
Using the peak of the dust continuum emission, 
pointing corrections were made.  
Pointing errors were estimated to be under $\sim$ 2".  
 A reduced- $\chi^{2}$ analysis of $Q$ and $U$ data \citep{dav11} was 
 performed to obtain 
 the final polarization map. 
The reduced $\chi^{2}$ was found to be 
1.87 $\pm$ 1.21 
when dividing the data into eight bins.  
The error bars for the polarization degree and magnetic field direction 
were inflated accordingly.  

\section{Results and Discussion}
\subsection{Measured Polarization and Magnetic Field} 
\begin{figure}[ht]
\begin{center}
\resizebox{8cm}{!}{\rotatebox{0}{\includegraphics{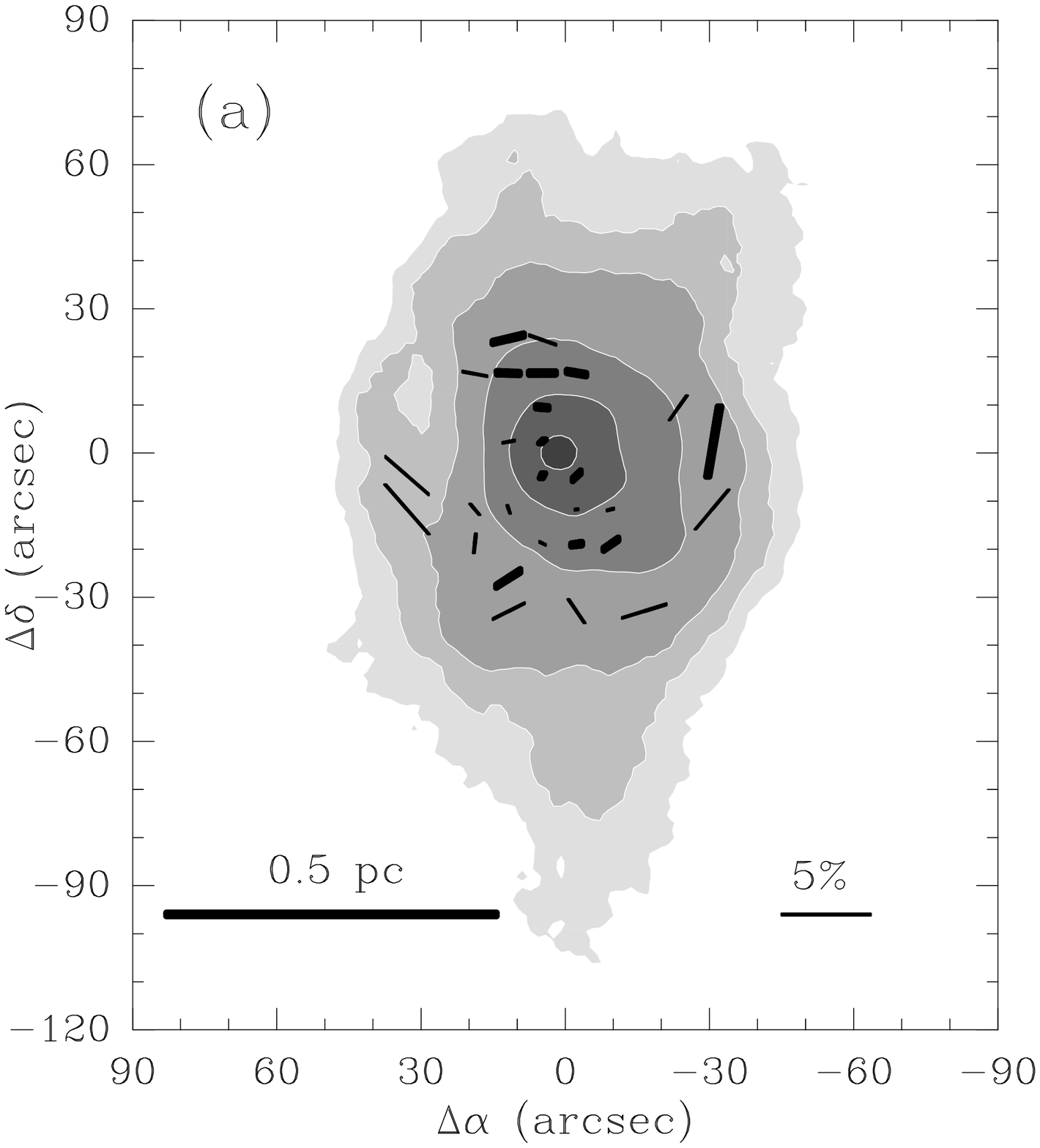}}}
\resizebox{8cm}{!}{\rotatebox{0}{\includegraphics{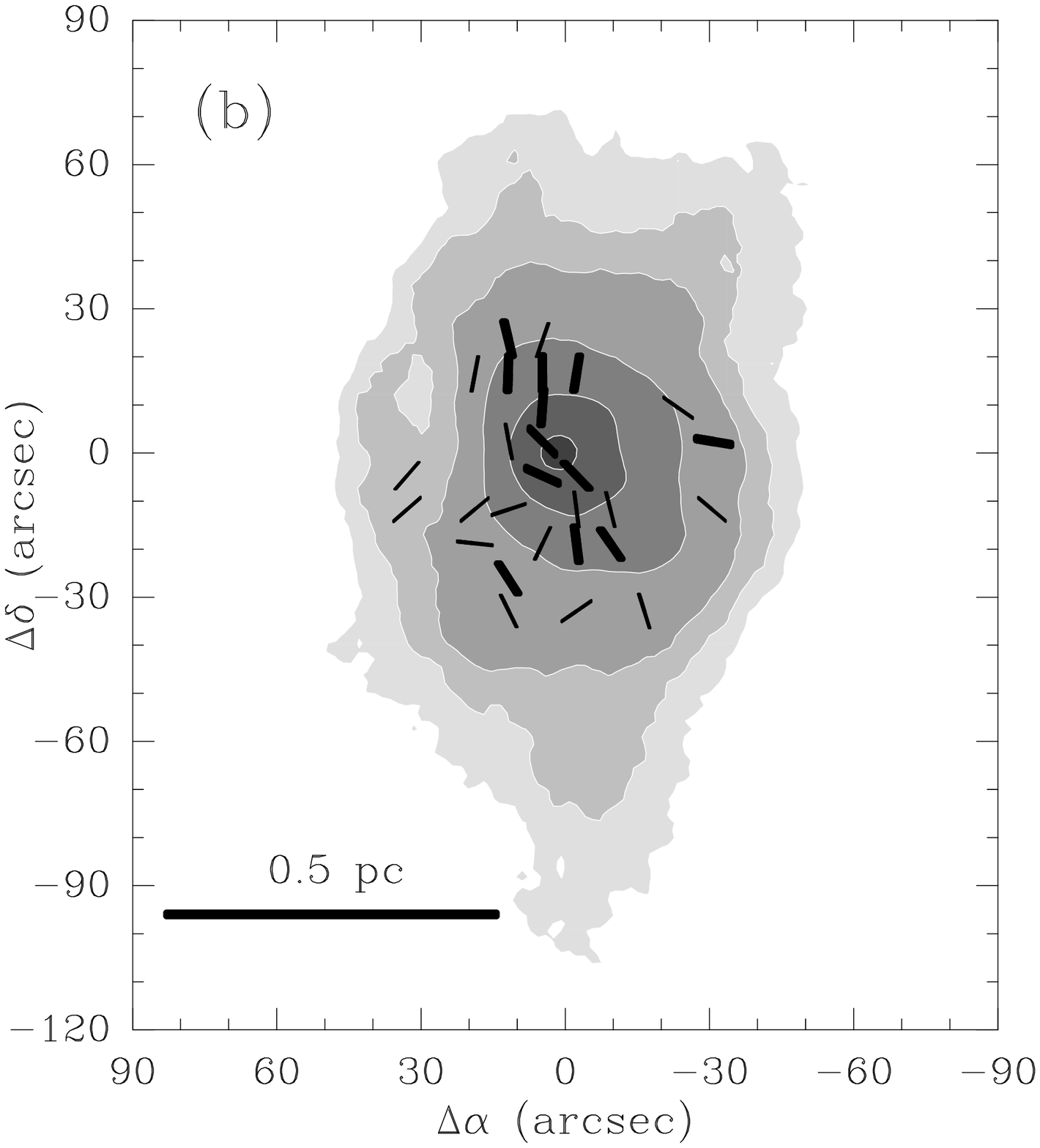}}}
\vspace{-1mm}
\caption{
Black bars overlaid on the 350$\mu$m dust continuum emission (grayscale) 
represent (a) the measured polarization E vectors at 350$\mu$m 
and (b) the measured magnetic field directions. 
The length of the black bars in (a) is set to be proportional to the polarization degree.  
The length corresponding to 5 \% polarization degree is shown at bottom right of the diagram.  
The contours represent the 350$\mu$m image obtained with SHARC II \citep{shi08} and   
are drawn at 2$\sigma$, 5$\sigma$, 9$\sigma$, 27$\sigma$, 81$\sigma$ and 243$\sigma$, 
where 2$\sigma$ corresponds to 200 mJy beam$^{-1}$.  
Thick bars and thin bars are vectors with signal-to-noise ratio 
between 2.5 and 6$\sigma$ and between 2 and 2.5$\sigma$, respectively.  
\label{polarivect}
}
\end{center}
\end{figure}

Figure \ref{polarivect} (a) shows the polarization vectors measured at 350 $\mu$m with SHARP.  
The degree of polarization in the envelope region is higher than that in the central region 
of the clump.  
Such a tendency, i.e., polarization hole, is observed towards other star forming regions such as Orion \citep{sch98} and W3 \citep{sch00}.  
\citet{sch98} argues that the decrease of polarization degree may be caused by temperature and/or optical depth effects.  
For the case of IRAS 20126+4104, 
the temperature structure appears to be complicated, as discussed in \citet{shi08}.  
The fact that vectors outside of the third contour from the peak of the dust continuum have larger polarization 
degrees compared to the vectors inside the third contour (Figure 1) suggests that the polarization hole may be caused by optical depth effects.  
\citet{mat09} reported the dust polarization of this object at 850 $\mu$m.  
The polarization vectors measured with SHARP are located in the central $r \sim$ 0.3 pc region  
of the massive dense clump, while the polarization vectors measured at 850$\mu$m trace 
a region outside of $r \sim$ 0.2 pc but within the $r \sim$ 0.5 pc region.  
Overall, the polarization vectors of Matthews et al. (2009) agree well with our 350 $\mu$m polarization vectors.  

The magnetic field directions are obtained by flipping the polarization vectors by 90\deg\ as plotted 
in Figure \ref{polarivect}  (b).  
Looking at the magnetic field directions along a north-south line passing through the center 
of the massive dense clump, 
one notices that many of these vectors (except for the central $\sim$ 0.1 pc region) roughly follow the north-south direction.  
On the other hand, for the east and west sides of the clump, many of the magnetic field directions tend to follow an east-west direction 
rather than a north-south direction.  
This tendency persists in the regions where 850$\mu$m polarization vectors are observed.  
Detailed comparison between the 350$\mu$m polarization vectors 
and the 850$\mu$m polarization vectors will be described elsewhere.  
The magnetic field changes its direction inside the infalling region, i.e., for radii below 0.1 pc. 


\subsection{Comparison with Simulation Results} 
To investigate the morphological evolution of magnetic field lines, we calculated
the evolution of a magnetized cloud using a three dimensional resistive
magnetohydrodynamics nested grid code with an isothermal equation of state
and sink cell treatment (Machida et al. 2005; 2011a).  Details are described in Kataoka et al. (in preparation).  
To conduct the simulation, we set the parameters so that they are close to the observed parameters 
of the massive dense clump IRAS 20126+4104.  

\begin{figure}[ht]
\begin{center}
\resizebox{9cm}{!}{\rotatebox{0}{\includegraphics{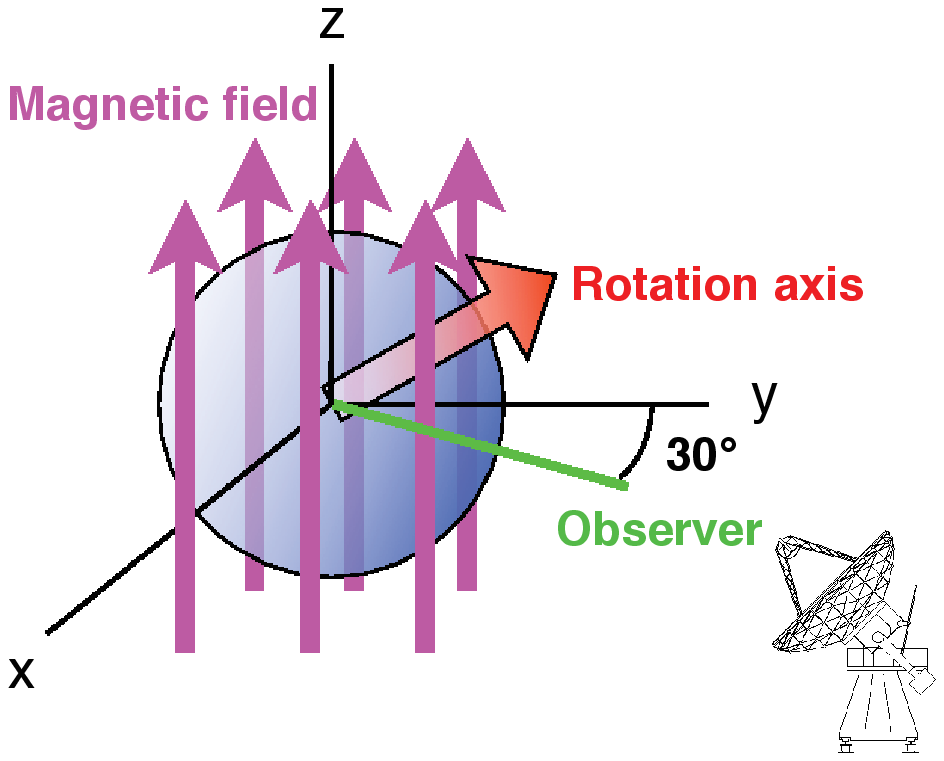}}}
\resizebox{12cm}{!}{\rotatebox{0}{\includegraphics{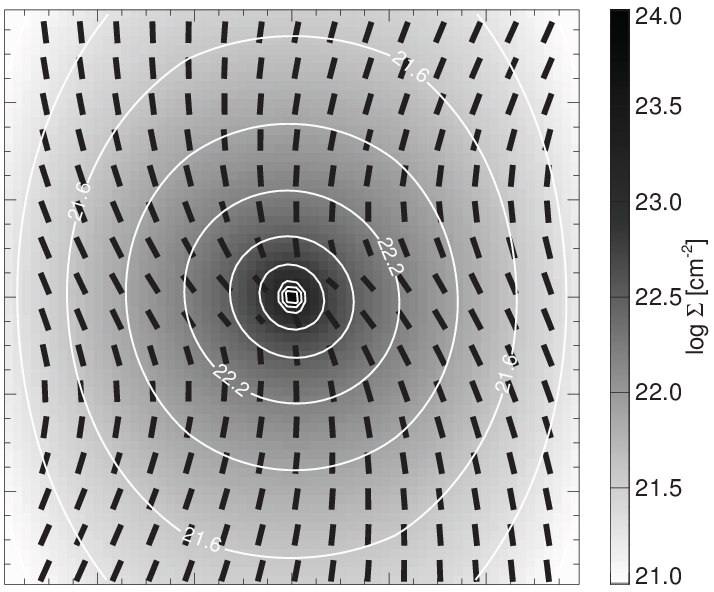}}}
\vspace{-1mm}
\caption{
Top: The schematic diagram summarizes the initial state of the simulation.  
The blue sphere represents the spherical cloud core, 
being penetrated by magnetic field (purple vectors) along the z axis.  
The rotation axis of the cloud core is inclined by 60$\degr$ from the z axis
on the y-z plane, as shown by the big red arrow.   
Note that observers view the cloud core from a direction 
that is inclined by 30$\degr$
from the y axis 
on the x-y plane.  
Bottom: The diagram shows the resultant magnetic field vectors (black bars) and the column density (grayscale and white contours) 
 calculated from the numerical simulation. 
White contours and grayscale represent the column density of the dense clump.
The protostar has a mass of 7.3M$_\odot$ and a large fraction of the
cloud mass remains as the infalling envelope.
The size of the diagram is 1.2 $\times 10^5$ by 1.2 $\times 10^5$ AU (=0.6 pc). 
\label{simulation}
}
\end{center}
\end{figure}

The schematic diagram in Figure \ref{simulation} shows the 
initial state, 
which is a spherical cloud core with a critical Bonnor-Ebert (BE)
density profile.  
The BE profile is characterized by two parameters,  the central number density
($n_{\rm c}$) and the isothermal temperature ($T_{\rm iso}$).
We adopt $n_{\rm c}=6\times10^3$\, cm$^{-3}$ with the density enhancement
factor of f=1.68 (Machida et al. 2011b) and $T_{\rm iso}$= 40\,K. 
Then, a uniform magnetic field ($B= 1.5\times10^{-5}$\,G) parallel to the z axis is
imposed on the whole computational domain.
The reason why we set the magnetic field direction to be parallel to the z axis is because 
the direction of the observed global magnetic field is almost north-south (see Section 3.4).  
In addition, rigid rotation ($\Omega_0= 1.1\times10^{-14}$\,s$^{-1}$
= 0.35 km s$^{-1}$ pc$^{-1}$ 
) is added to the
initial state, in which the rotation axis, defined by right-hand rule, is inclined with respect to the magnetic field 
(i.e., $z$-axis) at an angle of 
$60\degr$ on the y-z plane, as shown in the schematic diagram of Figure \ref{simulation}.   
The observer views the simulation from a point in the x-y plane.  This is shown in Figure 2, where it can be seen that 
the observer's line of sight is inclined by an angle of $30\degr$ with respect to the y axis.  
On the plane of the sky, 
the rotation axis becomes $P.A. \sim -40\degr$, 
close to the $P.A.= -35\degr$ rotation axis of the large-scale Keplerian disk (radius of 7400 AU  (0.037 pc); Cesaroni et al. 1999), 
which is situated at the center of the dense clump.  
With these assumptions, the initial cloud has a radius of $1.2\times10^5$\,AU (= 0.6 pc) and mass
of $78$\,M$_\odot$.
The initial cloud  has the energy ratios of E$_{thermal}$/E$_{gravity}$ = 0.5,
E$_{rotation}$/E$_{gravity}$ = 0.02 and E$_{magnetic}$/E$_{gravity}$ = 0.55,  where
E$_{thermal}$, E$_{rotation}$, E$_{magnetic}$, and E$_{gravity}$ are thermal, magnetic,
rotational and gravitational energy, respectively.
In the calculation, we assumed the protostar formation to occur when the central density exceeds the 
threshold number density 
$n_{\rm thr}=10^8$\,cm$^{-3}$  in the region of radius less than accretion radius, r$_{acc}$, 
where r$_{acc}$ = 64 AU is adopted as the sink radius (for details, see Machida
 et al. 2011a).
With this treatment, we calculated the cloud evolution until the gas accretion onto the protostar
or circumstellar disk almost halts.
To directly compare simulation results with our observations, we calculated polarization
of thermal dust emission using the formulation in Tomisaka (2011),
in which we assume that the whole region is optically thin and isothermal for simplicity.

The bottom diagram of Figure \ref{simulation} shows our simulation result, in which
magnetic field vectors and column density 7.1 $\times$ 10$^5$ yr after
the cloud collapse begins (or 8 $\times$ 10$^4$ yr after the protostar
formation) are plotted.  
The 30$\degr$ viewing angle (the top diagram of Figure \ref{simulation}) was chosen because 
it gives the best agreement between the observed and simulated magnetic field vectors.  
In particular, note that both Figure 1 (b) and Figure 2 show an S shape as one moves from north to south.  
At the epoch shown in the figure, the protostar has a mass of 7.3M$_\odot$ and a large fraction of the
cloud mass remains in the infalling envelope.
Note that radiative heating from the central protostar is ignored.

In a real high mass star forming region, one can expect in general that the rotation axis of a dense clump will not be aligned with the magnetic field. 
In this situation, simulations sometimes show 
the magnetic field morphology in an hourglass shape, 
but not always.  
Also, the observed 
shape of the magnetic field changes depending on the observer's viewing angle.  
For the simulation shown in Figure \ref{simulation}, 
the observed magnetic field vectors are aligned 
nearly in an S shape, but 
the morphology of the magnetic field vectors 
takes an hourglass shape 
if one observes the magnetic field along the x axis.  

In order to reproduce the observed magnetic field morphology, 
the effect of the rotation of the dense clump was found to be essential.  
This may imply that the rather fast rotation of the massive dense clump may have determined 
the magnetic field directions within the central regions of the massive dense clump,  
as well as the axis of the jet/Keplerian disk system associated with the central massive (proto)star.  
In our simulation, we do see the expected toroidal field generated
  by the rotating disk at the cloud's center.
  We cannot resolve it in Figure 2 because the scale
  of the toroidal field ($\sim 10^3\,$AU) is much smaller than the cloud
  scale ($\sim 10^5\,$AU).  Although the dominant components of the
  observed magnetic field presumably come from large scale structure,
  the innermost vectors of our magnetic field map may show
  a hint of the toroidal disk field.
 Although the dominant components of the observed 
  magnetic field come from large scale structure, 
  we might have detected a hint of toroidal field 
  generated by the rotating disk.

The configuration of the magnetic field lines is related to the
 rotating disk that has a size of $\sim 1000$\,AU. Since the sink
 radius adopted in this study (64 AU) is much smaller than the
 rotating disk, we can spatially resolve it. However, the resultant
 disk size may be dependent on the sink radius because we cannot
 resolve the early phase of the disk evolution.

\subsection{The Magnetic Field and Its Relationship with the Clump's Rotation, the Bipolar Outflow, and the Jet Directions} 
Based on the observed velocity gradient of the narrow line components, 
the rotation axis of the massive dense clump is roughly 140 $\pm$ 20$\degr$ in $P.A.$, 
where we again use the right-hand rule \citep{shi08}. 
On the other hand, the edge-on Keplerian disk of the massive (proto)star 
appears to rotate in almost the opposite sense to the rotation of the envelope of the massive dense clump 
whose effective radius is 0.56 pc (=1.15 $\times$ 10$^{5}$ AU) \citep{ces05,shi08}.  
\citet{mac06} discuss the cases when the rotation axis of a circumstellar disk is inclined 
with respect to the magnetic field axis.  
Under the conditions considered in their paper, i.e., a quiescent cloud core that forms low-mass stars, 
the counter rotation 
between cloud core and disk, as seen in IRAS 20126+4104, cannot occur.  
\citet{mac11b} discuss the importance of magnetic braking on the circumstellar disk formation in a strongly magnetized cloud.  
The kind of counter rotation observed in this object might happen when the angular momentum of a circumstellar disk is extremely efficiently transferred through magnetic braking.  

\begin{figure}[ht]
\begin{center}
\resizebox{6.5cm}{!}{\rotatebox{0}{\includegraphics{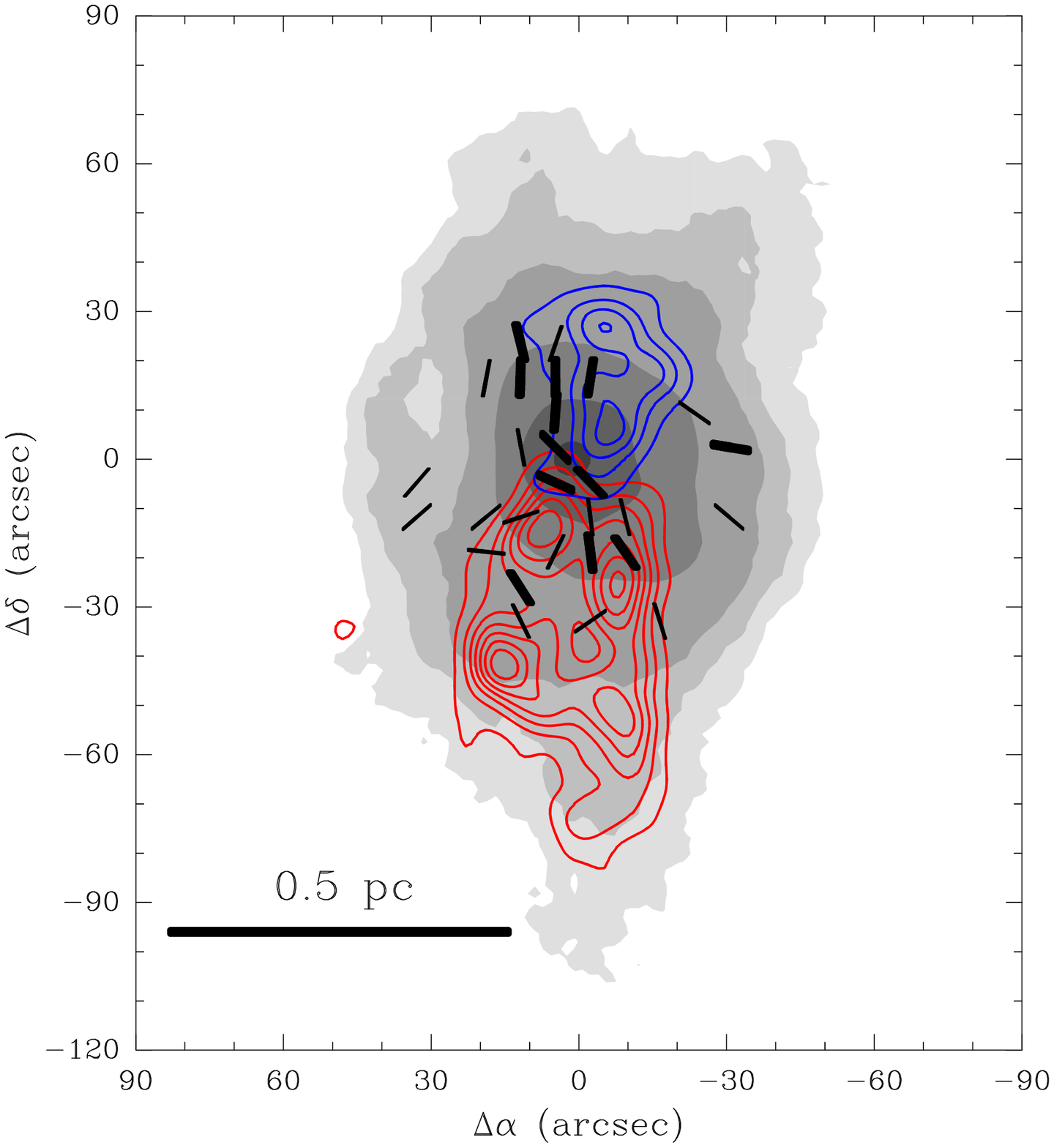}}}
\resizebox{6.5cm}{!}{\rotatebox{0}{\includegraphics{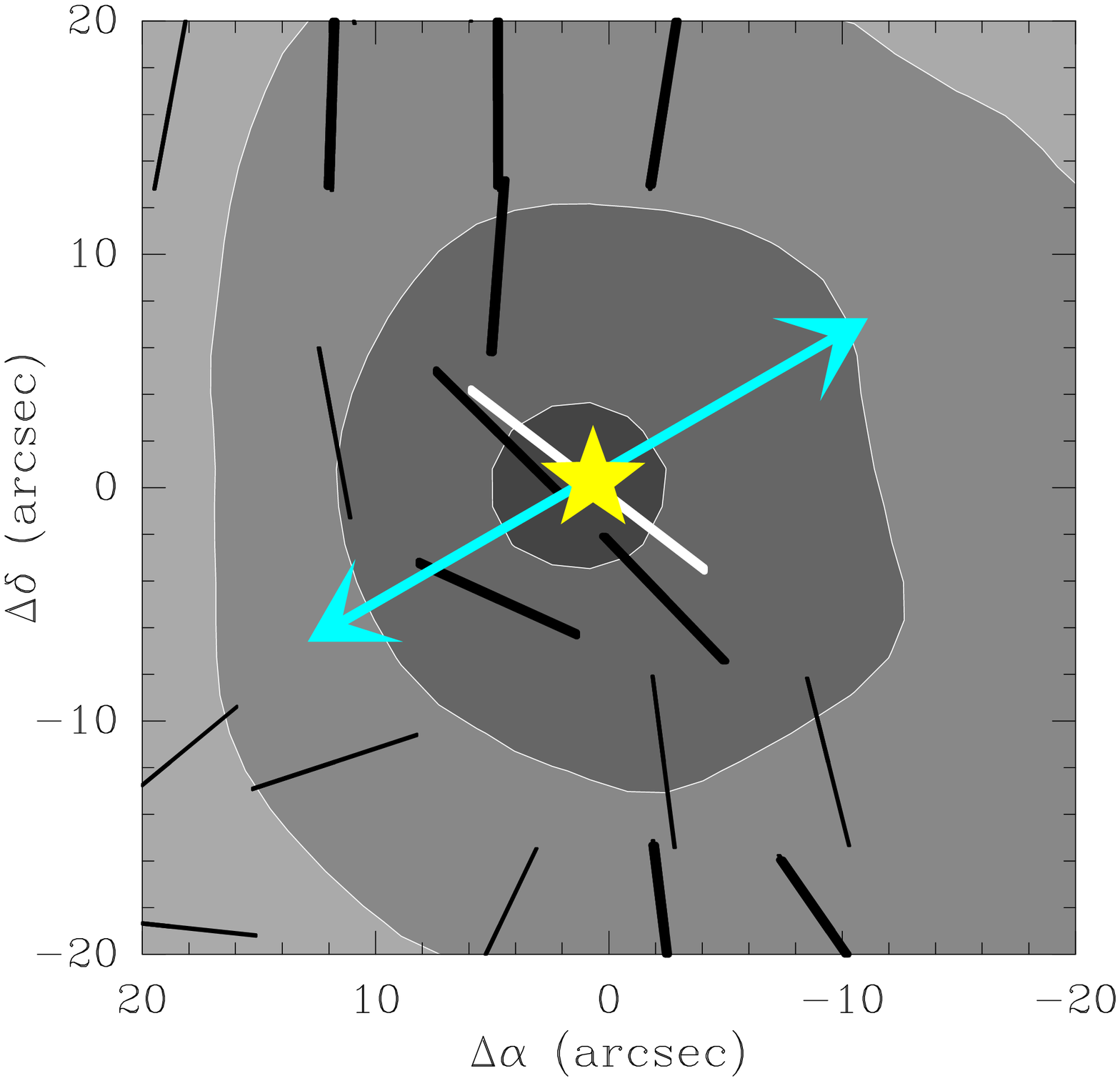}}}
\vspace{-1mm}
\caption{
Left:  
Molecular bipolar outflow lobes traced in the CO $J=1-0$ line (same as Figure 13 of \citet{shi08}) overlaid on the 
diagram of 
Figure 1 (b).  
Right: White line and light blue arrows 
represent the direction of the Keplerian disk ($P.A. \sim$ 53\deg ) and the direction of the jet ($P.A. \sim -$ 60\deg) associated with the massive (proto)star \citep{ces99}, marked with yellow star, overlaid on the diagram of Figure 1 (b), magnified in the central region.  
\label{outflowjet}}
\end{center}
\end{figure}

The two diagrams of Figure \ref{outflowjet}  show the cold bipolar outflow traced with CO $J=1-0$ and the directions of the jet/Keplerian disk plane, overlaid on the 
magnetic field directions and the 350 $\mu$m dust continuum map (same as Figure \ref{polarivect} (b)).  
The jet axis has $P.A. \sim -$60\deg. 
Comparing the bipolar outflow axis with the directions of the magnetic field in the massive dense clump, 
they are nearly parallel. 
For the warm bipolar outflow traced with CO $J=6-5$, the direction is very different (see Figure 12 of Shinnaga et al 2008), particularly the blue lobe is in a northwest-southeast direction with a smaller size than that of CO $1-0$.  
The direction of the CO $6-5$ outflow is close to the 
direction of the jet emanating from the massive (proto)star.  
%
The observed magnetic field directions near 
the Keplerian disk appear to lie parallel to the disk plane, which means at this location the fields are nearly 
perpendicular to the jet axis.

\citet{cia10} report numerical simulations of outflows in collapsing dense cores with misaligned rotation and magnetic field axes 
and find that a larger angle $\alpha$ between the rotation axis and the magnetic field direction leads to decreased efficiency 
of mass ejection via 
outflows.  The reason why the size of the CO $J=1-0$ bipolar outflow in IRAS 20126+4104 is larger than that of the 
CO $J=6-5$ outflow \citep{shi08} may be that 
$\alpha$ for the $1-0$ outflow is smaller than that of the $6-5$ outflow.  
\citet{cia10}  also found that the misalignment of magnetic field and rotation axes leads to jet precession.  
The apparent precession of the jet associated with the massive (proto)star in IRAS 20126+4104 thus 
may be explained by the misalignment of  magnetic field and rotation axis.  
An alternative explanation for precession is that the source at the clump center is binary, but observational studies carried out to date do not show evidence of binarity.


\subsection{The Magnetic Field in the Massive Dense Clump and  the Global Magnetic Field} 
\begin{figure}[ht]
\begin{center}
\plotone{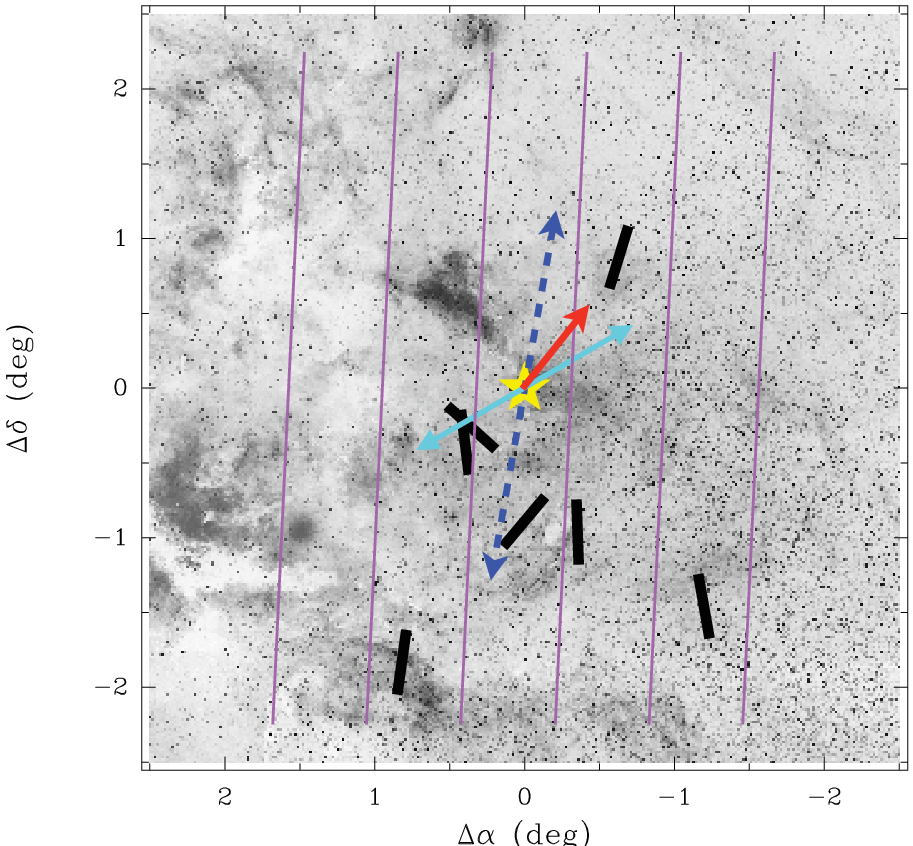}
\vspace{-9mm}
\caption{
Optical polarization vectors (black thick lines) are superposed on an optical image (DSS 0.5$\mu$m, gray scale)  
centered on IRAS 20126+4104.  The yellow star marks the position of the massive (proto)star. 
Purple thin lines represent the 
mean direction of the 
global magnetic field ($P.A. \sim -$ 3 deg).  
Blue thick dashed arrows show the direction of the CO $1-0$ bipolar outflow ($P.A. \sim-$ 10\deg).  
The red arrow and the light blue arrows   
represent the rotation axis of the large Keplerian disk ($P.A. \sim-$ 37\deg) and the directions of the jet ($P.A. \sim -$ 60\deg)
associated with the massive (proto)star, respectively.  
\label{globalB}
}
\end{center}
\end{figure}

It is of interest to compare the magnetic field direction in the massive dense clump and 
the global magnetic field surrounding the clump (e.g., Li et al. 2009).  
The global magnetic field direction around this object is estimated 
using the optical polarization data archive 
of Heiles (2000).  
The polarization data points that meet the following criteria were selected to trace   
the global magnetic field:  
(1) within $\pm$ 200 pc from the distance of the object (1500 pc) for the direction along the line of sight, 
and 
(2) within 2$\degr$ (roughly 50 pc in radius) from the center of the massive dense clump  on the plane of sky.  
The 
polarization data points and the mean direction of the global magnetic field are 
plotted in Figure \ref{globalB}.  
Note that 
four polarization data points that are within $\sim$ 0.8$\degr$ of the center of the nearby supernova remnant (SNR) G78.2+2.1 are omitted in this diagram 
because their polarization vectors are likely to be affected by the SNR.   
The threshold of 0.8$\degr$ was determined based on the radio continuum map of \citet{wed84}.
One sees that the global magnetic field is more or less aligned in a north-south direction, i.e., 
in the same direction as the 
CO $1-0$ bipolar outflow.  The massive dense clump is elongated in a north-south direction.  
The mean direction of the 
global magnetic field is  $P.A. \sim -3\degr$.  
The fact that the CO $1-0$ bipolar outflow is parallel to the global magnetic field suggests 
that magnetic field had a significant influence 
on the star formation process \citep{mat04}.   
However, the deeply embedded Keplerian disk and 
the jet show no such alignment with the global magnetic field.  
They seem to be aligned with the rotation axis of the dense clump.


\section{Summary and Future Work}   
In order to measure the magnetic field structure of the 
massive dense clump IRAS 20126+4104, 
polarized dust emission was measured at 350$\mu$m. 
The observed magnetic field 
is consistent with the magnetic field vectors of a simulation 
when magnetic field lines are inclined from the rotation axis of 
the dense clump. 
As an evidence that the magnetic field played a critical role on the formation of the massive (proto)star, 
the magnetic field directions within the massive 
dense clump 
are parallel to the CO $1-0$ bipolar outflow direction 
and appear to connect to the global magnetic field. 
Our observational data combined with the simulation results indicate that the rotation of the massive dense 
clump 
affects the magnetic field 
within the clump  as well as 
the orientation of the Keplerian disk/jet system.  
The counter-rotation feature observed between the envelope of the dense clump and the large rotating disk associated with the massive (proto)star 
might be a result of efficient 
angular momentum transfer via 
magnetic braking (e.g. Machida et al. 2011b).  
This issue should be further addressed in the future in order to fully understand 
the role of magnetic field 
in the star formation processes.  
It is important to measure the magnetic field strength within the massive dense clump 
using Zeeman effect 
in order to investigate the 
details of the evolution of the massive (proto)star and 
of the natal dense clump.

\acknowledgments
This research has been supported by NSF grant AST-0540882 and AST-0838261 to the CSO.  
SHARP has been supported by NSF grants AST 02-41356, AST 05-05230, 
and AST-0909030
to Northwestern University and grant AST 05-05124 to the University of Chicago. 
Numerical computations were carried out on NEC SX-9
and the general-purpose PC farm
at Center for Computational Astrophysics of
National Astronomical Observatory of Japan.
This work was supported by a Grants-in-Aid from MEXT (21740136,21244021).
Part of this work was carried out at the Jet Propulsion Laboratory, California Institute of Technology, under a contract with the National Aeronautics and Space Administration.
HS thanks Hua-bai Li, 
Roger Hildebrand, and Fumitaka Nakamura for helpful 
comments, and 
Shu-ichiro Inutsuka 
for valuable discussions. 



{\it Facilities:} \facility{CSO}.

\clearpage



\clearpage

\clearpage

\clearpage

\clearpage

\end{document}